\documentclass[aps,preprint]{revtex4}
\usepackage{amsfonts}
\usepackage{amssymb}
\usepackage{siunitx}
\usepackage{xcolor}
\usepackage{subcaption}
\usepackage{amsmath}
\usepackage{graphicx}
\usepackage[font={footnotesize,it}]{caption}
\usepackage{bigints}
\usepackage{latexsym}
\usepackage{enumitem}

\begin{document}

\title{Stationary, charged Zipoy-Voorhees metric from colliding wave spacetime}

\author{Mustafa Halilsoy}
 \email{mustafa.halilsoy@emu.edu.tr}

 \author{Mert Mangut \footnote{Corresponding author}}
 \email{mert.mangut@emu.edu.tr}

\author{Chia-Li Hsieh}
 \email{galise@gmail.com}

\affiliation{
 Department of Physics, Faculty of Arts and Sciences, Eastern Mediterranean University, Famagusta,
North Cyprus via Mersin 10, Turkey 
}

\begin{abstract}
Through the Ernst formalism we provide expression for a class of colliding Einstein-Maxwell (EM) metrics with cross polarization. Local isometry is imposed as a means to transform interaction region of the spacetime into stationary, charged Zipoy-Voorhees (ZV) metric in Schwarzschild coordinates. \textcolor{blue}{T}his is known as the Chandrasekhar-Xanthopoulos (CX) duality which maps the plane of double-null coordinates (with two spacelike cyclic coordinates) to the static/stationary spacetime. The ZV-metric is known to describe planetary/stellar objects with arbitrary distortion parameter. Asymptotic behaviour of the metric for practical use is provided.
\begin{center}
"The world may be seen in a grain of sand" - William Blake.
\end{center}

\end{abstract}

\keywords{Suggested keywords}
                              
\maketitle

\section{\label{sec:level1}Introduction}

Surely a grain of sand may not encompass the world , but Chandrasekhar-Xanthopoulos (CX)-duality between colliding plane waves (CPWs) spacetime and black holes (BHs), in analogy with the wave-particle duality \cite{1,12} may do the trick. This was the case of our recently found \cite{2} static, charged, ZV-metric \cite{3,4}. Such a process implicates a different kind of holography from the one between 4D inner region of a BH and 3D boundary at infinity. The AdS/CFT correspondence \cite{5} emerges as the resulting physics from the fact that the inner gravity is taken into account by the flat boundary of AdS, apt for quantum field theory. We note that our CX- duality, equivalently and under the title of Chandrasekhar-Ferrari-Xanthopoulos has been employed in \cite{18} to obtain the graviton scattering amplitudes at Planckian length. Our analysis in this study will be entirely classical.

In the CX-duality/holography, distinct from the familiar one of different dimensionalities, both spacetimes are in 4D and is valid also between CPW and non-BH spacetimes as we shall undertake in the present study. However, CPW geometry depends only on the two null coordinates $(u,v)$ or their useful combinations $(\eta(u,v),\mu(u,v))$ whereas it does not depend on the other coordinates $(x,y)$. The $(u,v)$, or $(\eta,\mu)-$ cell in the CPW spacetime which is curved  lies at a finite distance and maps to a solid metric in Schwarzschild (S) coordinates to represent arbitrarily shaped objects. In this duality electromagnetic (em) and gravitational wave parameters transform into charge and mass, respectively. In general relativity (GR) matter (energy) and charge derive from the bending and warping of spacetime which is manifested in the present study through the CX duality. In brief, a 'cell' of the $(\eta,\mu)$ coordinates with proper signature represents the 'whole', as expected from a hologram. Once the collision process is physically well-defined with proper boundary conditions their situation is reminiscent of the classical, i.e. non-quantum, version of the Breit-Wheeler formalism \cite{6} of colliding photons proposed long ago to create mass. In general relativity (GR) with colliding null sources such a process takes place naturally at the level of spacetimes and geometrically.

Organization of the paper is as follows. In section II by using the Ernst formalism we obtain a new metric representing colliding waves with cross-polarization. In section III we transform the metric into Schwarzschild coordinates to obtain the generalized version of the ZV-metric. Asymptotic expansions and closed timelike curves are discussed in section IV. (The details of expansions are tabulated in the Appendix). Conclusion and discussion in section V completes the paper.

\section{\label{sec:level2}New metric in the space of CPW$s$}

The spacetime of colliding Einstein-Maxwell (EM) waves is described by the line element \cite{1,7,10}

 \begin{equation}
ds^{2}= e^{\nu+\mu_{3}}\sqrt{\Delta}\left( \frac{d\eta^{2}}{\Delta}-\frac{d\mu^{2}}{\delta}  \right)-\sqrt{\Delta\delta}\left[ \chi dx^{2}+\frac{1}{\chi}\left(dy-q_{2} dx\right)^{2} \right]
\end{equation}

where $\chi$, $q_{2}$ and $\nu+\mu_{3}$ are metric functions depending only on $(\eta,\mu)$ with $\Delta=1-\eta^{2}$ and $\delta=1-\mu^{2}$. Note that the coordinates $(\eta,\mu)$ are functions of the null coordinates $(u,v)$ which are introduced for convenience. The Ernst equations \cite{8} are expressed in complex potentials $(Z,H)$ as 

\begin{equation}
\left(ReZ-\vert H \vert ^{2} \right)\nabla^{2}Z=(\nabla Z)^{2}- 2\bar{H}\nabla Z.\nabla H
\end{equation}

\begin{equation}
\left(ReZ-\vert H \vert^{2} \right)\nabla^{2}H=\nabla Z.\nabla H- 2\bar{H}(\nabla H)^{2}
\end{equation}

where the differential operators are defined appropriately on a basis manifold and a bar denotes complex conjugation. Here

\begin{equation}
Z=\Psi+\vert H \vert^{2}-i\Phi
\end{equation}

in which $\Psi$ and $\Phi$ are auxiliary complex potentials in the construction of the metric. For details one may consult \cite{1,9,10}. The vacuum case, i.e. $H=0,$ admits an integral \cite{9}

\begin{equation}
Z=\Psi-i\Phi=\frac{1}{D}(1-iq_{0}cosh2X)
\end{equation}

where $\vert q_{0} \vert\leq1$ is a constant that makes the metric stationary, which was denoted by $q_{0}=sin\alpha$ in \cite{9}. The function $D$ is given by

\begin{equation}
D=\sqrt{1+q_{0}^{2}}cosh2X-sinh2X
\end{equation}

where $X(\eta,\mu)$ is a harmonic function satisfying 

\begin{equation}
\left( \Delta X_{\eta} \right)_{\eta}-\left( \delta X_{\mu} \right)_{\mu}=0
\end{equation}

in which $()_{\eta}$ and $()_{\nu}$ stand for partial derivatives. Diverting now from \cite{9,10} we make the choice for the harmonic function 

\begin{equation}
e^{2X}=\left(\frac{1-\eta}{1+\eta} \right)^{\gamma}
\end{equation}

where $0<\gamma<\infty$ is an arbitrary distortion parameter that will be identified in the sequel as the ZV- parameter. Application of the CX-theorem  \cite{10} to the solution adapted in (8) serves to charge our cross-polarized metric in the space of CPWs. The metric reads 

\begin{equation}
ds^{2}=\frac{D\Omega^{2}}{4} \left[ \Delta^{\gamma^{2}}\left( \delta-\Delta \right)^{1-\gamma^{2}}\left( \frac{d\eta^{2}}{\Delta}-\frac{d\mu^{2}}{\delta}  \right)-\Delta\delta dx^{2} \right]-\frac{4}{D\Omega^{2}}\left(dy+q_{0}\gamma(1+p^{2})\mu dx \right)^{2}
\end{equation}

where $q=charge$, with $p^{2}+q^{2}=1$,

\begin{equation}
\Omega^{2}=\left(\frac{1-p}{D}+1+p\right)^{2}+\left(1-p \right)^{2}\left(\frac{q_{0}}{D}cosh2X \right)^{2}
\end{equation}

and $D$ is given in (6). The em complex potential $H$ is given by

\begin{equation}
H=q\frac{(1-p)(\Psi^{2}+\Phi^{2})+2p\Psi-(1-p)-2i\Phi}{(1-p)^{2}(\Psi^{2}+\Phi^{2})+2q^{2}\Psi-(1+p)^{2}}
\end{equation}

so that the vector potential $A_{\mu}=(0,A_{x},A_{y},0)$ has components

\begin{equation}
\begin{aligned}
&A_{y}=Re(H)\\
&A_{x}=Im(H)
\end{aligned}
\end{equation}

Let us remark that the metric (9) can be considered as a solution to the colliding EM waves provided the boundary conditions are satisfied without giving rise to current sources at the junctions. Since our aim here is not to search for the particular incoming wave profiles we shall ignore this part of the problem. Otherwise by going backward in time we can determine the waves that collide to give rise to the interaction region metric (9). In the next section we shall make use of the metric (9) of CPWs, and apply the transformation of CX to construct the corresponding stationary metric that interests us.

\section{\label{sec:level3}ZV- form of the metric in S-coordinates}

The crucial point now is to search for a transformation from $\{\mu,\eta,x,y\}$ coordinates of metric (9) into the S-coordinates $\{t,r,\theta,\varphi\}$. For this purpose we apply the transformation introduced first by CX \cite{1}. It should be added that the underlying transformation brings in extra constant coefficients, so that instead of an isometry it acts anti-homothetically. These constant terms, however, can be absorbed into a rescaling of coordinates. The transformation is

\begin{equation}
\begin{aligned}
&p\eta+1=\frac{r}{m}\\
&\mu=cos\theta\\
&x=\varphi\\
&y=\tau
\end{aligned}
\end{equation}

where $p$ is the em parameter and $m$ will be interpreted as the mass of the resulting source. We observe that while employing this transformatin the positive parameter $\gamma$ introduced in Eq. (8) is chosen to satisfy $(-1)^\gamma=-1$, to preserve the signature of the metric as $(-2)$. Although any real odd integer $\gamma$ does the job by resorting to complex analysis we expand our list of admissible values. Namely, from $e^{i(2k+1)\pi\gamma}=e^{i(2n+1)\pi}$ in which $k$ and $n$ are arbitrary integers, or zero, with the complex unit, $i=\sqrt{-1}$ gives $\gamma=\frac{2n+1}{2k+1}.$ The same result follows from the logarithmic branch consideration. The sequence of $\gamma's$ goes as $(...,\frac{1}{5}, \frac{1}{3}, \frac{3}{5}, \frac{5}{7}, 1, 3, 5, 7,...)$. The chose $(-1)^\gamma=1$, transforms our metric (9) into a metric with signature $(-,+,+,+)$ and the corresponding $\gamma$ will satisfy $\gamma=\frac{2n}{2k+1},$ where again $n$ and $k$ are arbitrary integers, including $k=0$. The $\gamma-$sequence for the latter choice goes as $(..., \frac{2}{5}, \frac{2}{3}, \frac{4}{3}, 2, 4, 6,... )$ which includes naturally the even integers. Given this information, metric (9) transforms into

\begin{equation}
ds^{2}=\frac{4}{D\Omega^{2}}\left(dt+q_{0}m\gamma p(1+p^{2})cos\theta d\varphi\right)^{2}-\frac{D\Omega^{2}}{4}\left\{\Sigma^{1-\gamma^{2}}\Delta_{0}^{\gamma^{2}}\left(\frac{dr^{2}}{\Delta_{0}}+r^{2}d\theta^{2} \right)+r^{2}\Delta_{0}sin^{2}\theta d\varphi^{2} \right\}
\end{equation}

where

\begin{equation}
\begin{aligned}
&\Delta_{0}=1-\frac{2m}{r}+\frac{m^{2}q^{2}}{r^{2}}\\
&\Sigma=1-\frac{2m}{r}+\frac{m^{2}}{r^{2}}\left( q^{2}+p^{2}sin^{2}\theta \right)\\
&D=\frac{1}{2\Delta_{0}^{\gamma}}\left\{(k-1)\left[1-\frac{m}{r}(1+p) \right]^{2\gamma}+(k+1)\left[1-\frac{m}{r}(1-p) \right]^{2\gamma}  \right\}\\
&\Omega^{2}=\left( \frac{1-p}{D}\right)^{2}\left(1+q_{0}^{2}cosh^{2}2X \right)-\frac{2q^{2}}{D}+(1+p)^{2}\\
&cosh2X=\frac{1}{2\Delta_{0}^{\gamma}}\left[\left(1-\frac{m}{r}(1+p) \right)^{2\gamma}+\left(1-\frac{m}{r}(1-p) \right)^{2\gamma}  \right]
\end{aligned}
\end{equation}

in which we have abbreviated  $k^2=1+q_{0}^2$. The em vector fields is given by $A_{\mu}=(A_{t},0,0,A_{\varphi})$. Now we have $A_{t}=mpRe(H)$ and $A_{\varphi}=Im(H)$, where

\begin{equation}
H=q\frac{(1-p)\left[1+(q_{0}cosh2X)^{2} \right]-D\left(2p+(1+p)D \right)+2iq_{0}Dcosh2X}{\left[(1-p)-(1+p)D) \right]^{2}+\left(q_{0}(1-p)cosh2X \right)^{2}}
\end{equation}

in which $D$ and $cosh2X$ are given in (15). In summary, our resulting solution has four parameters:

\bigskip
i) $m=$ mass

\bigskip
ii) $q=$ charge, with $q^{2}+p^{2}=1$

\bigskip
iii) $q_{0}=$ NUT-type parameter, with $k^{2}=1+q_{0}^{2}$ and $q_{0}^{2}\leq1$

\bigskip
iv) $\gamma=$ distortion (ZV) parameter, $0<\gamma<\infty$

\bigskip
Note that $\gamma>1$ is oblate and $\gamma<1$ is prolate. The limits of our metric (14) are as follow

\bigskip
a) $q_{0}=0=q\rightarrow$ ZV metric

\bigskip
b) $q_{0}\neq0=q\rightarrow$ NUT-type stationary metric, differing from \cite{11}. We note that, in Ref.\cite{11} quaternionic solution was used for the Ernst potential. In this paper we used a complex potential instead so that it slightly differs from  Ref.\cite{11} in choice of parameters.

\bigskip
c) $q_{0}=0\neq q$, $\gamma=1\rightarrow$ Reissner-Nordstrom metric

\bigskip
d) $q_{0}=0= q$, $\gamma=1\rightarrow$ Schwarzschild metric

\bigskip
e) $q_{0}=0\neq q$, $0<\gamma<\infty \rightarrow$ the metric in Ref \cite{2}

\bigskip

Due to the cross term $g_{t\varphi}\neq0$, our metric is not asymptotically flat. Similar to the NUT-metric \cite{13,14} it exhibits asymptotic flatness only in the plane $\theta=\pi/2$. We wish to add that the metric (14) is rather involved for an analytical treatment. Even the expression for $\sqrt{-g}$ in the wave equation analysis or Maxwell equations seems almost beyond analytical reach. The fact that all known limiting cases follow from the solution given is shown above. Further, in principle, multiplying the choice for $e^{2X}$ in (8) by an exponential factor involving a quadrupole term, for instance, provides additional contributions to the metric.

\section{\label{sec:level3}Asymptotic form of the metric, closed time-like curves}

\subsection{Metric functions for $r\rightarrow\infty$}

For large $r$ we introduce a new radial variable $R=r_{0}r,$ where

\begin{equation}
r_{0}=(k+1)p^2+k-1
\end{equation}

and express all metric functions in the following expansions up to the order $1/R^3$. We have

\begin{equation}
\begin{aligned}
&g_{tt}\approx 1+\sum_{n=1}^{3}a_{n} R^{-n}   \\
&g_{t\varphi}\approx  b_{0}\left(1+\sum_{n=1}^{3}b_{n} R^{-n}\right) \\
&g_{rr}\approx c_{0}\left(1+\sum_{n=1}^{3}c_{n} R^{-n}\right)\\
&g_{\theta\theta}\approx \sum_{n=-2}^{3}d_{n} R^{-n}\\
&g_{\varphi\varphi}\approx \sum_{n=-2}^{3}e_{n} R^{-n}
\end{aligned}
\end{equation}

where all coefficients are tabulated in the Appendix $(A1-A5)$. Similarly, the vector potential $A_{\mu}=(A_{t},0,0,A_{\varphi})$ is expanded to the same order as follows

\begin{equation}
\begin{aligned}
&A_{t}\approx \alpha_{0}\left(-1+\sum_{n=1}^{3}\alpha_{n} R^{-n}\right)   \\
&A_{\varphi}\approx  \beta_{0}\left(1+\sum_{n=1}^{3}\beta_{n} R^{-n}\right)
\end{aligned}
\end{equation}

The coefficients $\alpha_{n}$, $\beta_{n}$ can be found in the Appendix, $A6$ and $A7$. It can easily be checked that the limit for $k=1$, $p=1$, our metric reduces to the $ZV-$ metric, and for $\gamma=1$ it gives the Schwarzhild metric.

\subsection{Possible existence of closed, time-like curves}

As in many other stationary metrics closed time-like curves can be searched in the well-known manner a la' Gödel \cite{16}. The metric is glued isometrically projected to the $(t,\varphi)$ sector, by choosing $r=const.$, and $\theta=const.$, in the following manner.

\begin{equation}
ds^2=e^{2\Psi}\left( dt-\omega d\varphi\right)^2-e^{-2\Psi}r^2\Delta_{0}sin^2\theta d\varphi^2
\end{equation}

Here we have abbreviated $e^{2\Psi}=\frac{4}{\Omega^2D}$ with $\omega=-q_0mp\gamma(1+p^2)cos\theta$. We choose $t=-a\varphi$, where $a$, is a small, positive parameter i.e $a<<1$. so that $\varphi=0$ and $\varphi=2\pi$ are identified and the two Killing vectors $\partial_t$ and $\partial_\varphi$ are proportional. Upon substitution into the metric and demanding that

\begin{equation}
\left(a+\omega \right)^2-r^2\Delta_{0}e^{-4\Psi}sin^2\theta>0
\end{equation}

will provide a closed time-like curve with a non-causal region such that the end point $-2\pi a$ precedes the initial point $t=0$. Since we have adopted $r=const.$, $\theta=const.$ and $0<\gamma<\infty$ there is certain range for both variables to make (21) satisfied. In Fig.1 we plot $e^{-2\Psi}$ versus $\gamma$ for small $r$ and the region in between the two curves is such a possible region. We recall that a special case of stationary metric, known as NUT-Curzon without charge \cite{12} was also shown long ago to admit similar closed time-like curves. See also \cite{19} with references cited therein. Let us add that the inequality (21) can be satisfied also in the absence of charge. That corresponds to a chargeless, non-asymptotically flat ZV metric where $q_0$ parameter makes it stationary. The topology of our metric is cylindrical and any causality violation assumed will not lead to a torus typs topology which incorporates holes. 

\begin{figure}[h]
\centering
\includegraphics{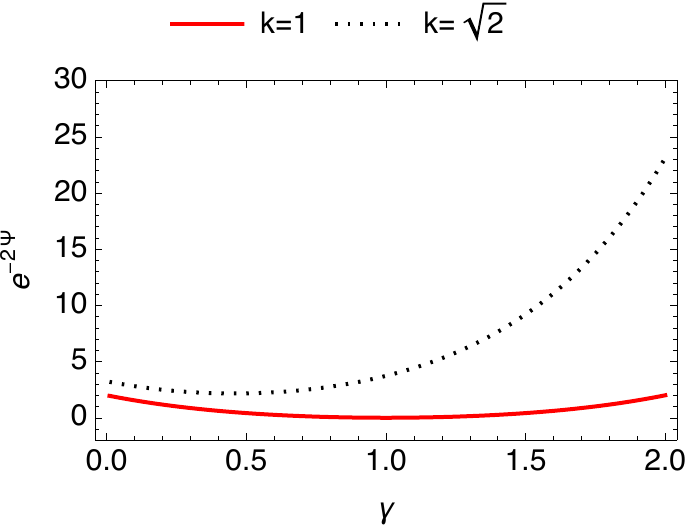}
\caption{The minimum $(k=1)$ and maximum spin $(k=\sqrt{2})$ are considered in the plot of $e^{-2\Psi}$ versus $\gamma$. In between the two curves we can make our inequality to be satisfied for small $r$ and reasonable choice of $\gamma.$ Addition of a constant to the metric function $\omega$ can also be satisfied.}
\end{figure}

\section{\label{sec:level3}Conclusion and discussion}

The aim of this paper was twofold. Firstly to extend the CX-duality to cover local isometry between CPW spacetimes with non-BH ones such as the topological ZV-class. Secondly, by employing this duality to obtain the stationary, NUT-like charged extension of the static ZV-metrics. (See also \cite{12}). The four-parametric spacetime obtained represents non-spherical objects with the distortion parameter $0<\gamma<\infty$ which is specified as a discrete set. Although the resulting metric is more realistic than the spherical ones, i.e. $\gamma=1$, for an analytical treatment it is extremely involved. To this end we rely on the CX-theorem \cite{9,10} valid in the $(\eta,\mu,x,y)$ coordinates followed by the holographic transformation into the $(t,r,\theta,\varphi)$ coordinates. Obviously the factor that makes the spacetime so complicated is the distortion parameter $\gamma\neq1$. All limiting cases, such as charged ZV, ZV, Reissner-Nordstrom (RN) and Schwarzschild are perfectly satisfied. We considered briefly going backward in time in our new universe. It is needless to add that this concerns our material world alone since conservation of history should disallow the same for biological time. Finally, from curiosity we wish to add the following comment. To resurrect the Breit-Wheeler \cite{6} idea, albeit at a classical level, we ask: Can we obtain any given physical metric from the collision of massless gauge fields, such as em waves superposed with gravitational waves? In case the answer is affirmative then, "Let there be light" must be revised as "Let there be collision of light" to create the material world. This is one possible answer to the quest, what makes the mass and charge? Based on the CX-duality many examples are available to justify such a view. We remind that any spacetime admits a plane wave as a limit \cite{15} which may be considered as incoming state prior to a collision.

\section*{Acknowledgements}
We thank the referee for much valuable comments. \\

\textbf{Appendix A. }\\

Asymptotic expansion coefficients for metric functions $(A1-A5)$ and electromagnetic potentials  $(A6,A7)$.

\begin{equation} \tag{A1}
\begin{aligned}
&a_{1}=-4 mp^2\gamma \;\;\;\;\;\;\;\;\;\;\;\;\;\;\;\;\;\;\;\;\;\;\;\;\;\;\;\;\;\;\;\;\;\;\;\;\;\;\;\;\;\;\;\;\;\;\;\;\;\;\;\;\;\;\;\;\;\;\;\;\;\;\;\;\;\;\;\;\;\;\;\;\;\;\;\;\;\;\;\;\;\;\;\;\;\;\;\;\;\;\;\;\;\;\;\;\;\;\;\;\;\;\;\;\;\;\\
&a_{2}=- \left(2 p^2(k+1) +\gamma k \left(k \left(p^2+1\right)^2+p^4-1\right)+2 k-8\gamma p^2-2\right) 2 m^2p^2\gamma\\
&a_{3}=8m^3 p^4 \gamma ^3 \left(5 k^2 \left(p^2+1\right)^2+4 k \left(p^4-1\right)-\left(p^2+22\right) p^2-1\right)\\
&-\frac{4}{3}\gamma m^3p^2 \left(p^2+3\right) r_{0}^{2} -4 \gamma^2 m^3 p^2 \left(k \left(k \left(p^2+1\right)^2+p^4-1\right)-8 p^2\right)
\end{aligned}
\end{equation}

\begin{equation}\tag{A2}
\begin{aligned}
&b_{0}=2mp\gamma c(\theta)\\
&c(\theta)=(p^2+1)\sqrt{k^2-1}\cos (\theta )\\
&b_{1}=-4m\gamma p^2\\
&b_{2}=-2m^2p^2 \gamma\left( 2p^2 (k+1)+\gamma k \left(k \left(p^2+1\right)^2+p^4-1\right)+2 k-8\gamma p^2-2\right) \\
&b_{3}=-\frac{8}{3}p^4m^3\gamma ^3 \left(-5 k^2 \left(p^2+1\right)^2-4 k \left(p^4-1\right)+p^4+22 p^2+1\right)\\
&+\frac{4}{3}r_{0}p^2m^3\gamma\left(\left(p^2+3\right) \left((k+1) p^2+k-1\right)+3\gamma \left(k \left(k \left(p^2+1\right)^2+p^4-1\right)-8 p^2\right)\right)
\end{aligned}
\end{equation}

\begin{equation}\tag{A3}
\begin{aligned}
&c_{0}=\frac{1}{4} r_{0}^2\\
&c_{1}= 4m\gamma p^2\\
&c_{2}=\frac{1}{2}m^2 p^2r_{0}\left(k\left(3 \gamma ^2+1\right)  \left(p^2+1\right)-\left(\gamma ^2-1\right) p^2-1\right)\;\;\;\;\;\;\;\;\;\;\;\;\;\;\;\;\;\;\;\;\;\;\;\;\;\;\;\;\;\;\;\;\;\;\;\;\;\;\;\;\;\;\;\;\;\;\;\;\;\;\;\;\;\;\;\;\\
&+\frac{1}{2} m^2 p^2 r_{0}\left(\gamma  (\gamma +8)+r_{0}\left(\gamma ^2-1\right)  \cos (2 \theta ) \right) \\
&c_{3}=-2m^3 p^2 r_{0}^2\left(\gamma ^2-1\right)  \left(k p^2+k+2 \gamma  p^2+p^2-1\right)\sin ^2(\theta )\\
&-\frac{4}{3}m^3 p^2  r_{0}^2\gamma\left(3 \gamma  k+p^2 \left(2 \gamma ^2+3 \gamma  k+1\right)+3\right)
\end{aligned}
\end{equation}

\begin{equation}\tag{A4}
\begin{aligned}
&d_{-2}=\frac{1}{4}\\
&d_{-1}=-\frac{1}{2}m \left(k p^2+k- 2 \gamma p^2+p^2-1\right) \\
&d_{0}=+\frac{1}{8} m^2p^2r_{0}^2 \left(\gamma ^2-1\right)\cos (2 \theta ) \\
&+\frac{1}{8} m^2r_{0}\left(k \left(p^2+1\right) \left(\left(3 \gamma ^2-1\right) p^2+2\right)-\left(\gamma ^2+1\right) p^4+((\gamma -8) \gamma +3) p^2-2\right)\\
&d_{1}= \frac{1}{6}(3 \cos (2 \theta )+1) \left(\gamma ^2-1\right) \gamma  m^3 p^4r_{0}^{2}\\
&d_{2}=\frac{r_{0}^{3}}{196}\left(\left(\gamma ^2-1\right) \gamma  m^4 p^4 \left(32-\gamma\left( (7 k-9) p^2+7 k+9\right)+12\cos (2 \theta )\left(\gamma +3\gamma k+p^2\gamma (3 k-1) +8\right)\right)\right)\\
&+ \frac{3}{196}\left(\gamma ^2-1\right) \gamma^2  m^4 p^4 r_{0}^{4} \cos (4 \theta )\\
&d_{3}= \frac{1}{480}r_{0}^{4} \gamma  \left(\gamma ^2-1\right) m^5 p^4\left(80-5 \gamma  (7 k+9)-p^2 (\gamma  (6 \gamma +35 k-45)+16)\right)\\
&+\frac{1}{24} r_{0}^{4} \gamma  \left(\gamma ^2-1\right) m^5 p^4\left(3 (\gamma +3 \gamma  k+4)+p^2 (\gamma  (2 \gamma +9 k-3)+4)\right)\cos (2 \theta ) \\
&+\frac{1}{32} r_{0}^{4}\gamma  \left(\gamma ^2-1\right) m^5 p^4 \gamma  \left(k p^2+k+2 \gamma  p^2+p^2-1\right)\cos (4 \theta ) 
\end{aligned}
\end{equation}

\begin{equation}\tag{A5}
\begin{aligned}
&e_{-2}=\frac{1}{4r_{0}}\sin ^2(\theta ) \\
&e_{-1}=-\frac{1}{2}(-1 + k + p^2 + k p^2 - 2 p^2 \gamma) sin^{2}(\theta)\\
&e_{0}=- (-1 + k^2)(1 + p^2)^2 m^2 p^2  \gamma^2 cos^{2}(\theta)\\
&+\frac{1}{2}m^2r_{0}\left(k \left(p^2+1\right) \left(\left(2 \gamma ^2-1\right) p^2+1\right)-p^2 \left(4 \gamma +p^2-2\right)-1\right)\sin ^2(\theta ) \\
&e_{1}=\frac{2}{3}m^3 p^4\gamma \left( 6\gamma^2\left(k^2-1\right) \left(p^2+1\right)^2 \cos ^2(\theta )+\left(\gamma ^2-1\right) r_{0}^{2}\sin ^2(\theta )\right) \\
&e_{2}=12 \gamma ^2 \left(k^2-1\right) \left(p^2+1\right)^2\left(2 (k+1) p^2+\gamma  k \left(k \left(p^2+1\right)^2+p^4-1\right)+2 k-8 \gamma  p^2-2\right) \cos ^2(\theta ) \\
&+r_{0}^{3}\left(\gamma ^2-1\right) \left(\gamma  k \left(p^2+1\right)+4\right) \sin ^2(\theta )\\
&e_{3}=\frac{16}{3} \gamma^5  m^5 p^6 \left(k^2-1\right) \left(p^2+1\right)^2  \left(-5 k^2 \left(p^2+1\right)^2-4 k \left(p^4-1\right)+p^4+22 p^2+1\right)\cos ^2(\theta ) \\
&+\frac{24}{3} \gamma^4  m^5 p^4 r_{0}\left(k^2-1\right) \left(p^2+1\right)^2 \left(k \left(k \left(p^2+1\right)^2+p^4-1\right)-8 p^2\right)\cos ^2(\theta )\\
&+\frac{8}{3} \gamma^3  m^5 p^4 r_{0}^{2}\left(k^2-1\right) \left(p^2+1\right)^2 \cos ^2(\theta )\\
&+\frac{2}{15} \gamma  m^5 p^4r_{0}^{4} \left(\gamma ^2-1\right) \left(5 (\gamma  k+2)+p^2 \left(2 \gamma ^2+5 \gamma  k+2\right)\right)\sin ^2(\theta ) 
\end{aligned}
\end{equation}

\begin{equation}\tag{A6}
\begin{aligned}
&\alpha_{0}=\frac{p(k+1)}{(k+1) p^2+k-1}q\\
&\alpha_{1}=\frac{2 \gamma  m }{k+1} \left(k \left(p^2-1\right)+p^2+1\right)\\
&\alpha_{2}=\frac{2 \gamma  m^2 p}{k+1} \left(2 \gamma  p^2 \left(k^2+(k-2) (k+1) p^2+k-2\right)+k \left((k+2) p^4-k+2\right)+p^4-1\right)\\
&\alpha_{3}=\frac{2 \gamma  m^3r_{0} }{3(k+1)}\left(r_{0} \left(p^2+3\right)\left(k \left(p^2-1\right)+p^2+1\right)+12 \gamma  p^2\left(k^2+(k-2) (k+1) p^2+k-2\right)\right)\\
&-\frac{4 \gamma^3p^6 m^3}{3}\left((k (2 k+15)-23)+(k+1) (2 k-1) p^2\right)\\
&+\frac{4 \gamma^3p^2 m^3(k-1)}{3(k+1)}\left((-1 + k) (1 + 2 k) + (-23 + k (-15 + 2 k)) p^2\right)
\end{aligned}
\end{equation}

\begin{equation}\tag{A7}
\begin{aligned}
&\beta_{0}=\frac{\sqrt{k^2-1}}{(k+1) p^2+k-1}q\\
&\beta_{1}=4 \gamma  m p^2\\
&\beta_{2}=2 \gamma   m^2 p^2 \left(\gamma +k \left(p^2+1\right) \left(\gamma  \left(p^2-1\right)-2\right)+p^2 \left(\gamma  \left(p^2+6\right)-2\right)+2\right)\;\;\;\;\;\;\;\;\;\;\;\;\;\;\;\;\;\;\;\;\;\;\;\;\;\;\;\;\;\;\;\;\;\;\;\;\;\;\;\;\;\;\;\;\;\;\;\;\;\;\;\;\;\;\;\;\\
&\beta_{3}=-\frac{4 r_{0}\gamma m^3 p^2}{3}\left(\left(p^2+3\right)r_{0}-3 \gamma  \left(k \left(p^4-1\right)+p^4+6 p^2+1\right)\right)\\
&+\frac{16\gamma^3 m^3 p^4(k-2)}{3}  \left(2 \left(k-2\right) p^2+k^2+ (k+1) p^4+1\right)
\end{aligned}
\end{equation}

\end{document}